\documentclass{svproc}

\usepackage{url}

\usepackage{amsmath}
\usepackage{amssymb}
\usepackage{graphicx}
\usepackage[font=small,skip=0pt]{caption}
\usepackage[hidelinks]{hyperref}

\newcommand{\setDec}{D} %
\newcommand{\safeLevel}{h} %

\newcommand{\R}{\mathbb{R}} %
\newcommand{\Rp}{\mathbb{R}_{>0}} %
\newcommand{\Np}{\mathbb{N}_{>0}} %

\newcommand{\Pb}{\mathbb{P}} %

\begin{document}
\mainmatter              %
\title{Safety in safe Bayesian optimization and its ramifications for control}
\titlerunning{Safety in Safe Bayesian Optimization}  %
\author{Christian Fiedler$^\ast$\inst{1} \and Johanna Menn$^\ast$\inst{1} \and Sebastian Trimpe\inst{1}}
\authorrunning{Fiedler, Menn, Trimpe} %
\tocauthor{Christian Fiedler, Johanna Menn, and Sebastian Trimpe}
\institute{Institute for Data Science in Mechanical Engineering (DSME), RWTH Aachen University, Aachen, Germany\\
\email{\{fiedler,johanna.menn,trimpe\}@dsme.rwth-aachen.de}\\
($^\ast$ Equal contribution)}

\maketitle              %

\begin{abstract}
A recurring and important task in control engineering is parameter tuning under constraints, which conceptually amounts to optimization of a blackbox function accessible only through noisy evaluations.
For example, in control practice parameters of a pre-designed controller are often tuned online in feedback with a plant, and only safe parameter values should be tried, avoiding for example instability.
Recently, machine learning methods have been deployed for this important problem, in particular, Bayesian optimization (BO).
To handle safety constraints, algorithms from safe BO have been utilized, especially SafeOpt-type algorithms, which enjoy considerable popularity in learning-based control, robotics, and adjacent fields.
However, we identify two significant obstacles to practical safety. First, SafeOpt-type algorithms rely on quantitative uncertainty bounds, and most implementations replace these by theoretically unsupported heuristics.
Second, the theoretically valid uncertainty bounds crucially depend on a quantity - the reproducing kernel Hilbert space norm of the target function - that at present is impossible to reliably bound using established prior engineering knowledge.
By careful numerical experiments we show that these issues can indeed cause safety violations.
To overcome these problems, we propose Lipschitz-only Safe Bayesian Optimization (LoSBO), a safe BO algorithm that relies only on a known Lipschitz bound for its safety.
Furthermore, we propose a variant (LoS-GP-UCB) that avoids gridding of the search space and is therefore applicable even for moderately high-dimensional problems.
\end{abstract}
This extended abstract, which has been presented as a poster at the Symposium on Systems Theory in Data and Optimization (SysDO) 2024, disseminates results from the journal paper  \cite{fiedler2024losbo}.
The content of all sections is adapted and all plots and results are taken verbatim from \cite{fiedler2024losbo}.

\section{Introduction} \label{sec:intro}
A recurring and very important task in science, engineering and business is the optimization of a blackbox function that is only accessible through noisy evaluations.
Frequently, each function evaluation is costly, and hence the optimization process should use as few trials as possible.
In control engineering, this situation occurs for example when tuning the parameters of a pre-designed controller in feedback with the plant.
In this situation, the search space consists of all parameter values for the controller, the unknown function is a given performance criterion, 
and each function evaluation consists in running the closed-loop system with a given controller setting for a certain amount of time.
Since running such an evaluation is time-consuming and causes the plant to wear and tear, as few function evaluations as possible should be used for the tuning process. Furthermore, since every real-world plant is subject to (random) disturbances and measurement errors, the true map from controller parameters to the performance measure is only approximately available, i.e., we have to deal with noisy function evaluations.
The optimization of a blackbox function accessible only through noisy evaluations can be tackled with Bayesian optimization (BO) \cite{garnett2023bayesian,Shahriari2016}.
This machine learning technique maintains an internal model of the unknown function based on past function evaluations and prior knowledge, like smoothness or periodicity, and in each optimization step, this internal model is used to determine the next input to be evaluated.
The latter step is commonly implemented by maximizing an internal acquisition function, which in turn is derived from the current model \cite{garnett2023bayesian}.
BO has a long history and is by now a very established technique with many real-world use cases and industrial-scale software implementations \cite{garnett2023bayesian,Shahriari2016}. 
Recently, BO has been successfully used for tuning controller parameters online \cite{marco2016automatic,neumannbrosig2019data}.
However, this use case of BO has an additional important complication.
Since here a function evaluation corresponds to running the controller in feedback with the actual physical plant, only controller parameters should be tried that are safe to run on the physical plant \cite{vonrohr2024local}.
For example, only parameters that correspond to a stabilizing controller should be suggested by the optimization algorithm.
The issue also appears in BO applications in robotics \cite{baumann2021gosafe,marco2021robot}.
What makes this problem very challenging is that, in general, it is not known whether a given parameter value is safe to run. 
To address this issue, safe BO techniques have been developed \cite{kim2020safe}.
Let $\setDec$ be the set over which the optimization is to be performed.
In the case of controller tuning, this corresponds to the set of all possible controller parameters.
Let $f_\ast: \setDec \rightarrow \R$ be the target function that is to be optimized, which is unknown in general.
In the case of controller tuning, this corresponds to the map from controller parameters to a performance measure of interest, e.g., the infinite-horizon cost of an LQR controller.
A generic BO algorithm suggests in every iteration $t\in\Np$ of the optimization process an input $x_t\in\setDec$, on which the target function $f_\ast$ is queried, and a noisy evaluation $y_t = f_\ast(x_t)+\eta_t$ is received by the algorithm.
The noise $\eta_1,\eta_2,\ldots$ is commonly modeled as independent and identically distributed (i.i.d.) zero-mean random variables, and often additional assumptions like subgaussianity are imposed.
BO algorithms typically maintain an internal model $\mathcal{M}_t$ based on the observed data $(x_1,y_1),\ldots,(x_t,y_t)$ up to time $t$, starting with an initial model $\mathcal{M}_0$ encoding prior knowledge about $f_\ast$.
At time step $t\in\Np$, the next input is typically determined by the optimization problem $\max_{x \in \setDec} \alpha_t(x; \mathcal{M}_{t-1})$,
where $\alpha_t(\cdot; \mathcal{M}_{t-1}): \setDec \rightarrow \R$ is called an acquisition function,
and must only depend on the internal model $\mathcal{M}_{t-1}$, but not on the unknown target function $f_\ast$.
For a comprehensive overview of acquisition functions, their theoretical properties, and implementation issues, we refer to \cite{garnett2023bayesian}.

In applications like controller tuning, safety constraints have to be taken care of. We focus on the setting of one safety constraint on the performance measure, which corresponds to the requirement that the controller performance must not be below a certain minimum level. This setting can be extended to multiple safety functions, which are independent of the performance measure, a detailed discussion can be found in \cite{kim2020safe}. 
Formally, there exists a set of safe inputs $S\subseteq D$, which is typically unknown, and the optimization algorithm should only suggest inputs $x \in S$.
In safe BO, the set $S$ is usually encoded as $S = \{ x \in \setDec \mid f(x) \geq \safeLevel \}$
where $\safeLevel$ is a known safety threshold. A safe BO algorithm decides upon the next query input using the constrained optimization problem
\begin{equation}
     \max_{x \in S_t(\mathcal{M}_{t-1})}\alpha_t(x; \mathcal{M}_{t-1}),
\end{equation}
where $S_t(\mathcal{M}_{t-1}) \subseteq \setDec$ is an estimate of the set $S$ based only on the internal model $\mathcal{M}_{t-1}$, but not on $S$.
It is clear that without any additional assumptions, no algorithm can succeed in solving this problem.
Most commonly the knowledge of some nonempty $S_0\subseteq S$ is required, and the optimization algorithm starts its exploration on this set.

To summarize, given initial knowledge about the target function $f_\ast$ in the form of an initial model $\mathcal{M}_0$, as well as a known set $S_0\subseteq S$, a safe BO algorithm should optimize $f_\ast$ with few function evaluations, and only querying $f_\ast$ with inputs $x \in S$.
This problem setting has been introduced in the seminal work \cite{sui2015safe}, which also proposed and analyzed the SafeOpt algorithm.
Many variants of the latter algorithm have been studied in a variety of settings \cite{baumann2021gosafe,berkenkamp2023bayesian,berkenkamp2016safe,sui2015safe,sui2018stagewise},
including controller tuning \cite{berkenkamp2016safe,berkenkamp2023bayesian}, %
and safe robot learning \cite{baumann2021gosafe}.
Furthermore, this class of algorithms comes with rigorous safety and performance guarantees \cite{baumann2021gosafe,berkenkamp2023bayesian,sui2015safe,sui2018stagewise}.
However, in \cite{fiedler2024losbo}, we identified several subtle issues related to safety in real-world applications of this class of algorithms.
In this extended abstract, we report on these findings from the perspective of control applications, as well as our proposed solutions.
\section{The problem with safety in SafeOpt-type algorithms} \label{sec:safetyProblem}
\begin{figure}[htbp!]
    \hfill
    \includegraphics[width=0.49\textwidth]{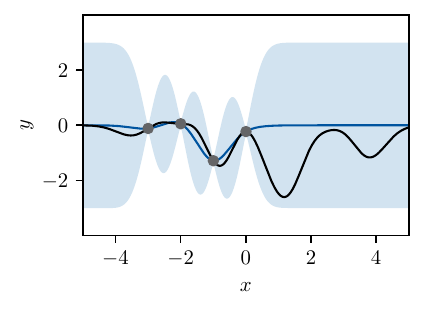}
    \hfill
    \includegraphics[width=0.49\textwidth]{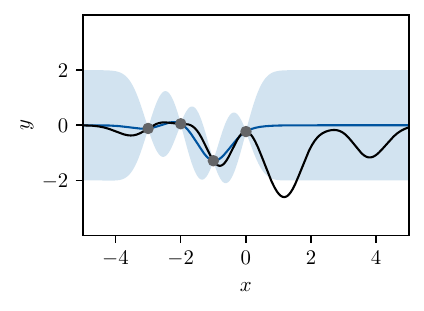}
    \hfill
    \caption{Illustration of the required GP error bounds. Consider a fixed ground truth (solid black line), of which only finitely many samples are known (black dots). Applying GP regression leads to a posterior GP fully described by the posterior mean (solid blue line) and the posterior variance, 
    from which a high-probability uncertainty set can be derived (shaded blue).
    Left: The ground truth is completely contained in the uncertainty set. 
    Right: The ground truth violates the uncertainty bound around $x=1$.
    Figure from \cite{fiedler2024losbo}.}
    \label{fig:illustrationUniformBoundsGPR}
\end{figure}

SafeOpt-type algorithms, starting with \cite{sui2015safe}, use as their internal model Gaussian process (GP) regression \cite{Rasmussen2006}, a nonparametric Bayesian regression method.
The GP prior, essentially a probability distribution over functions, corresponds to the initial model $\mathcal{M}_0$, encoding prior knowledge about the target function (and analogously about the safety functions).
Assuming independent Gaussian additive noise, GP regression allows updating the model with observational data $(x_1,y_1),\ldots,(x_t,y_t)$ in closed form using only linear algebra routines, resulting in the posterior model $\mathcal{M}_t$.
The latter provides with the posterior mean $\mu_t$ a nominal prediction of the target function,
and with the posterior variance $\sigma_t^2$ a measure of uncertainty \cite{Rasmussen2006}.
The basic idea of SafeOpt-type algorithms is to compute from $\mathcal{M}_{t}$ a lower bound $\ell_t$ on the target function $f_\ast$ that holds with high probability. 
More formally, we need $\Pb[f_\ast(x) \geq \ell_t(x) \: \forall x \in \setDec,\: t \geq 1]\geq 1-\delta$, where $\delta\in(0,1)$ is chosen by the user. 
It is possible to derive a sequence of scaling factors $(\beta_t)_t\subseteq\Rp$ such that $\Pb\left[f_\ast(x) \in [\mu_t(x)-\beta_t\sigma_t(x), \mu_t(x)+\beta_t\sigma_t(x)] \: \forall x \in \setDec,\: t \geq 1\right]\geq 1-\delta$, cf. \cite[Theorem~6]{Srinivas2010},\cite{abbasi2012online}, and \cite[Theorem~2]{Chowdhury2017}.
This is illustrated in Figure \ref{fig:illustrationUniformBoundsGPR} (left).
Therefore, one can choose $\ell_t(x)=\mu_t(x)-\beta_t\sigma_t(x)$, which has been indeed proposed for SafeOpt-type algorithms \cite{sui2015safe}.
However, the bound \cite[Theorem~6]{Srinivas2010} suffers from large numerical constants,
and as the improved bound \cite[Theorem~2]{Chowdhury2017} can be difficult to evaluate.
This might explain why to the best of our knowledge, \emph{all} previous implementations of SafeOpt-type algorithms replace theoretically justified bounds by heuristics, usually $\beta_t \equiv 2$ or some other constant.
As a result, \emph{all theoretical safety guarantees} are lost.
In \cite{fiedler2024losbo}, we demonstrate with numerical experiments that even in relatively benign settings, such heuristics can lead to significant bound violations, cf. Figure \ref{fig:illustrationUniformBoundsGPR} (right) for an illustration.
\vspace{-5mm} 
In \cite{fiedler2021aaai}, a variant of the bound \cite[Theorem~2]{Chowdhury2017} was proposed that is easy to evaluate, and numerical experiments showed that the resulting uncertainty sets are only slightly larger than common heuristics.
Motivated by these developments, we proposed in \cite{fiedler2024losbo} to use such a bound in the actual implementation of the SafeOpt algorithm (called Real-$\beta$-SafeOpt to distinguish this from implementations relying on heuristics), which results in an algorithm that retains all theoretical guarantees.
Numerical experiments in \cite{fiedler2024losbo} show that this algorithm performs well, and indeed maintains safety.
However, the bound from \cite{fiedler2021aaai} (using improvements from \cite{abbasi2012online}) requires that the data is generated from a target function $f_\ast$ contained in the reproducing kernel Hilbert space (RKHS) generated by the covariance function used in GP regression, and that an upper bound on the corresponding norm of $f_\ast$ is known.
A RKHS is a Hilbert space of functions in which function evaluation is continuous w.r.t. the norm.
These function spaces are very popular in machine learning, statistics, and numerics \cite{SC08}, and assuming that $f_\ast$ is contained in such an RKHS is in most cases a harmless assumption, cf. the discussion in \cite{fiedler2024losbo}.
In contrast to this, assuming that an upper bound on the RKHS norm is known is a severe limitation.
In particular, while the RKHS norm is a well-understood object, to the best of our knowledge, it is at present not possible to derive an upper bound using typical engineering prior knowledge, cf. \cite{fiedler2024losbo} for an extensive discussion.
We would like to stress that a valid upper norm is not only necessary for the theoretical analysis of SafeOpt-type algorithms, \emph{but for the proper working of the algorithm itself}.
Indeed, numerical experiments in \cite{fiedler2024losbo} confirm that an invalid upper norm bound leads to significant safety violations.
To summarize, as long as the safety of the algorithm relies on a quantitative upper bound on the RKHS norm of the target function, it is at present not possible to ensure that SafeOpt-type algorithms are safe in practice.
\vspace{-5mm} 
\begin{figure}[htpb!]
    \centering
    \includegraphics[width=0.8\textwidth]{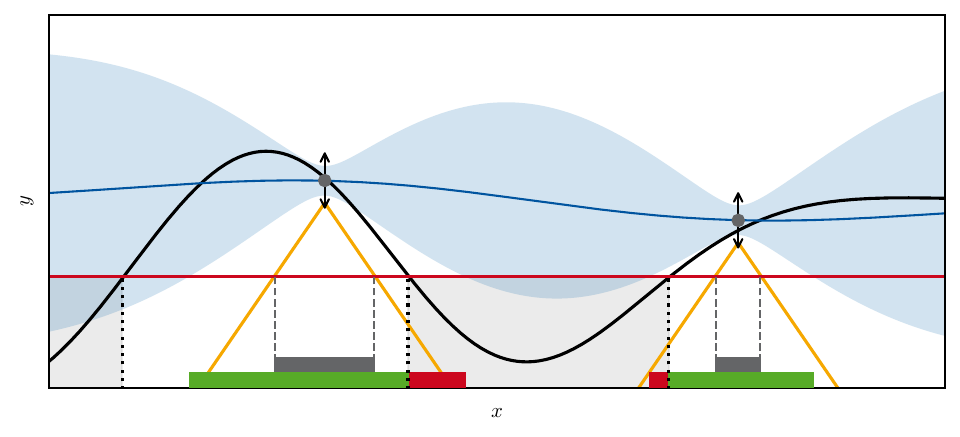}
    \caption{Illustration of LosBO being safe, while a safe set based on invalid uncertainty bounds leads to potential safety violations.
    The safe set of LoSBO (gray set) is determined by the constant $E$ (gray arrow) and the Lipschitz cone (orange). 
    The GP mean and the confidence bounds are illustrated in blue. 
    The points in the safe set given by the lower confidence bound are green if they are safe and red if they are unsafe. Figure from \cite{fiedler2024losbo}.}
    \label{fig:losboIllustration}
\end{figure}
\section{LoSBO} \label{sec:losbo}
\begin{figure}[htpb!]
     \includegraphics[width=\textwidth]{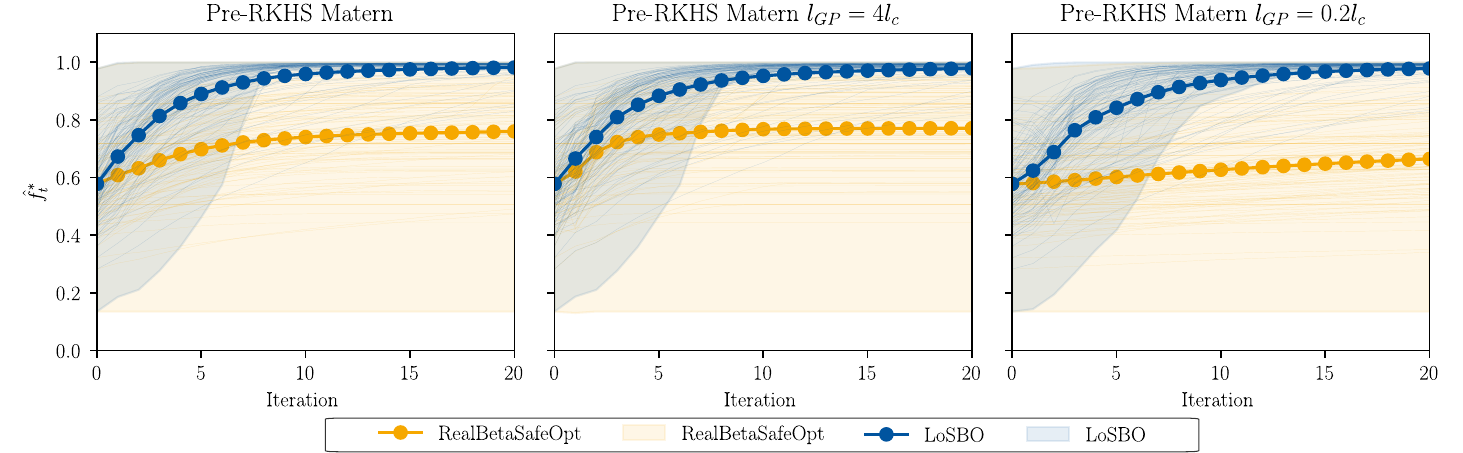}

  \vspace{2mm}
    \caption{Comparison of LosBO and Real-$\beta$-SafeOpt in a well-specified and misspecified setting. Thick solid lines are the means over all functions and repetitions, thin solid lines are the means over all repetitions for each individual function, shaded area corresponds to one standard deviation over all runs.
    Figure from \cite{fiedler2024losbo}.}
    \label{fig:experiments:losboAndRealBetaSafeOptWellspecified}
\end{figure}

To address the serious safety issue described above, we propose a variant of SafeOpt in \cite{fiedler2024losbo} called Lipschitz-only Safe Bayesian optimization (LoSBO).
The SafeOpt algorithm assumes that the target function is $L$-Lipschitz continuous, and uses this to extrapolate the uncertainty bounds described above.
However, a known Lipschitz bound is a common assumption in systems and control, cf. \cite{fiedler2024losbo} for a discussion of this point, and if one assumes bounded noise, as is also done frequently in systems and control, then it is possible to derive quantitative uncertainty bounds without any additional assumptions.
This is well-known in systems identification, and used for example in nonlinear set membership identification \cite{milanese2004set}.
The idea of LoSBO is to separate the exploration (i.e., optimization) and safety mechanisms, and using only a known Lipschitz bound and bounded noise for safety.
This is illustrated in Figure \ref{fig:losboIllustration}. These assumptions have clear interpretations, are natural in many applications, and are established in domains like control engineering. However, the applicability of the proposed algorithm clearly hinges on these assumptions, and they have to be judge on a case-by-case base by practitioners. 
For example, a Lipschitz constant can be derived from physical prior knowledge, or from high-fidelity simulations.
In particular, the scalings $\beta_t$ are now proper hyperparameters of the BO algorithm that can be freely tuned without compromising safety. 
Using extensive numerical experiments on synthetic functions, we show in \cite{fiedler2024losbo} that LoSBO performs very favorably compared with other SafeOpt-type algorithms, while having practically meaningful safety guarantees.
Some selected experimental results are presented in Figure \ref{fig:experiments:losboAndRealBetaSafeOptWellspecified}.
\section{LoS-GP-UCB} \label{sec:losgpucb}
\begin{figure}[!htbp] \centering
    \includegraphics[width=0.4\textwidth]{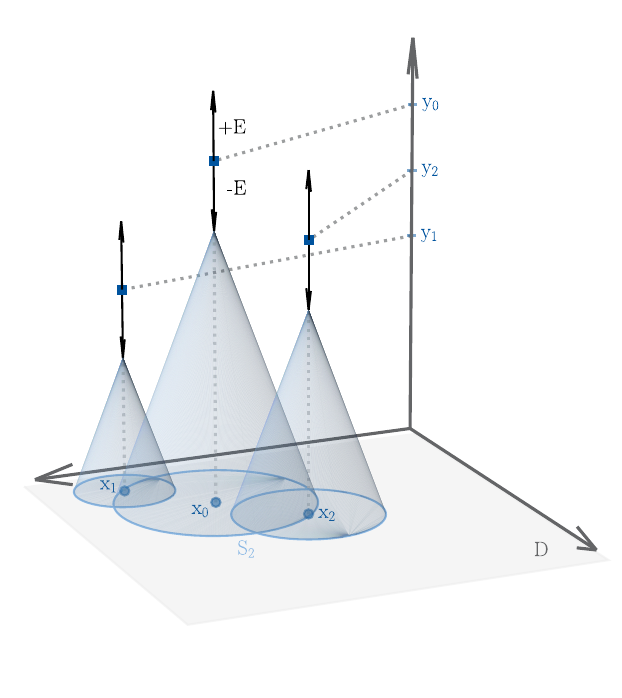}
    \includegraphics[width=0.4\textwidth]{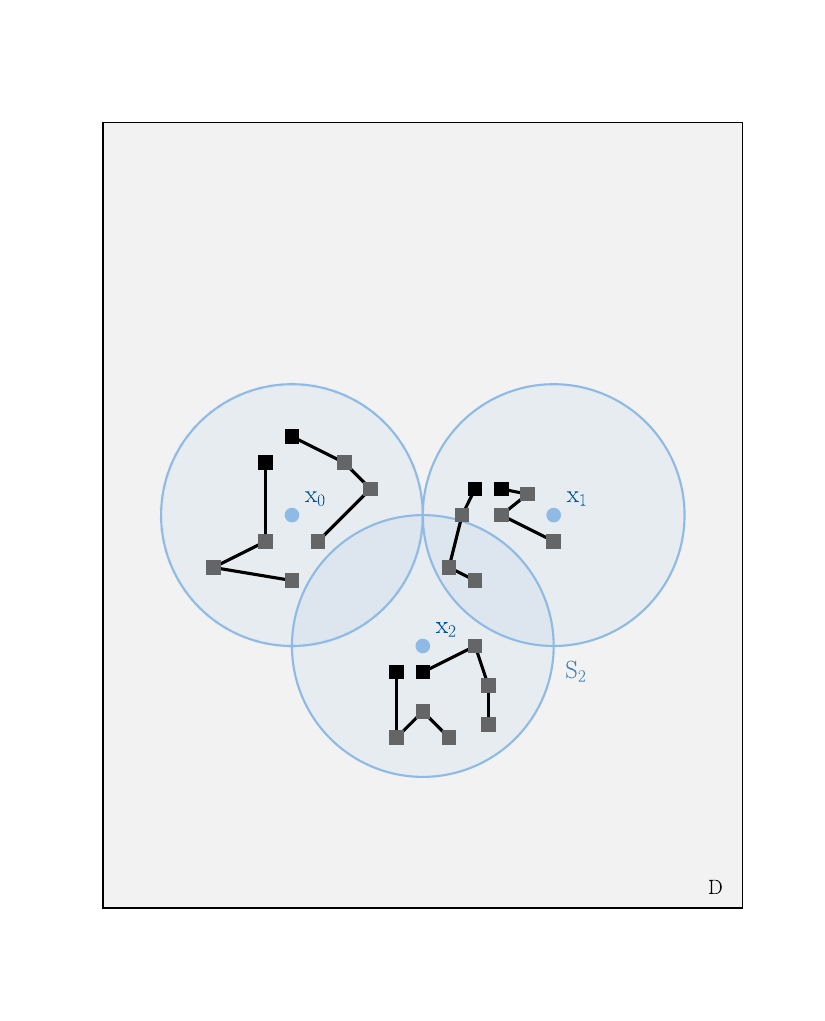}
    \caption{Illustration of one iteration of LoS-GP-UCB. 
    Figure from \cite{fiedler2024losbo}.}
    \label{fig:illustrationLoS-GP-UCB}
\end{figure}

Like most SafeOpt-type algorithms, LoSBO requires a discrete search space.
Since in practice, for example in controller tuning, the parameter space is continuous, gridding needs to be applied, which becomes prohibitive even in moderate dimensions.
To overcome this issue, we propose a variant of LoSBO, Lipschitz-only safe GP Upper Confidence Bound (LoS-GP-UCB), that avoids gridding.
The idea is to apply the safety mechanism to the popular GP-UCB BO algorithm \cite{Srinivas2010}, 
and then use a local search method with random multistarts to optimize the acquisition function,
which is the standard in modern software implementations like BOtorch \cite{balandat2020botorch}. 
This is illustrated in Figure \ref{fig:illustrationLoS-GP-UCB}.
We investigated the performance of LoS-GP-UCB using extensive numerical experiments with synthetic and benchmark functions, revealing excellent performance while maintaining safety, based only on clearly interpretable and reasonable assumptions.

\section{Conclusion and outlook} \label{sec:conclusion}
The iterative optimization of an unknown function under safety constraints is an important problem in science and engineering, including in particular automatic tuning of controller parameters on a plant.
In \cite{fiedler2024losbo}, we investigated two significant practical safety issues of the popular class of SafeOpt-type algorithms, and proposed the LoSBO algorithm as a solution, and LoS-GP-UCB as a variant suitable for even moderately high-dimensional problems.
Ongoing work is concerned with applications to tuning an automotive tracking controller, handling multiple safety constraints \cite{menn2024}, and variants suitable for high dimensional problems.

\vspace{1em}
\textbf{Acknowledgements} This work was performed in part within the Helmholtz School for Data Science in Life, Earth and Energy (HDS-LEE),
and funded in part by the German Federal Ministry for Economic Affairs and Climate Action (BMWK) through the project EEMotion. Computations were performed with computing resources granted by RWTH Aachen University under project rwth1459.

\bibliographystyle{spmpsci} 
\bibliography{cleanrefs} 

\end{document}